\documentclass[journal,letterpaper]{IEEEtran_nssmic}

\usepackage{graphicx}  

\usepackage{url}       
\begin{document}
%
\title{The Offline Software Framework of the Pierre Auger Observatory}
%
%
\author{S. Argir{\`o},
        S.L.C. Barroso,
        J. Gonzalez,
        L. Nellen,
        T. Paul,
        T.A. Porter,\\
        L. Prado Jr.,
        M. Roth,
        R. Ulrich and
        D. Veberi{\v c}
\thanks{The work of J. Gonzalez and T. Paul was funded in 
part by the National Science Foundation.  The work of T.A. Porter
was funded in part by the US Department of Energy.
}
\thanks{S. Argir{\'o} is with the INFN and University of Torino, 
S.L.C. Barroso is with the Centro Brasileiro de Pesquisas F{\'i}sicas,
J.Gonzalez and T. Paul are with Northeastern University,
L. Nellen is with the Universidad Nacional Aut{\'o}noma de M{\'e}xico,
T. Porter is with Louisiana State University,
L. Prado Jr. is with the University of Campinas,
M. Roth and R. Ulrich are with the Forschungzentrum Karlsruhe,
and D. Veberi{\v c} is with Nova Gorica Polytechnic.
}}
\maketitle

\begin{abstract}
The Pierre Auger Observatory is designed to unveil the nature and the
origins of the highest energy cosmic rays. The large and geographically
dispersed collaboration of physicists and the wide-ranging collection of simulation
and reconstruction tasks pose some special challenges for the offline
analysis software.
We have designed and implemented a general purpose framework which allows 
collaborators to contribute algorithms and sequencing instructions to build up the
variety of applications they require.  The framework includes
machinery to manage these user codes, to organize the abundance of
user-contributed configuration files, to facilitate multi-format file
handling, and to provide access to event and time-dependent detector
information which can reside in various data sources.  A number of
utilities are also provided, including a novel geometry package which
allows manipulation of abstract geometrical objects independent of
coordinate system choice. The framework is implemented in C++, 
and takes advantage of object oriented design and common open source
tools, while keeping the user side simple enough for C++ novices to
learn in a reasonable time.
The distribution system incorporates unit and acceptance
testing in order to support rapid development of both the core
framework and contributed user code.  
\end{abstract}


\section{Introduction}
%
%
%
%
\PARstart{T}{he} offline software framework of the 
Pierre Auger Observatory~\cite{Abraham:2004dt} 
provides an infrastructure to support a variety of distinct 
computational tasks
necessary to analyze data gathered by the observatory.
The observatory itself is designed to measure the
extensive air showers produced by the highest
energy cosmic rays ($> 10^{19}$~eV) with the goal 
of discovering their origins and composition.  
Two different techniques are used to detect
air showers.  Firstly, a collection of telescopes
is used to sense the fluorescence light produced
by excitation of nitrogen induced by the cascade
of particles in the atmosphere.
This method can be used only when the sky is moonless and dark, 
and thus has roughly a 10\% duty cycle.
The second method uses an array of 
detectors on the ground to sample particle 
densities as the air shower arrives at the Earth's surface.
Each surface detector consists of a tank containing 12 tons of
purified water instrumented with photomultiplier tubes to detect
the Cherenkov light produced when particles pass through
it.  The surface detector has a 100\% duty cycle.
A subsample of air showers detected by both instruments,
dubbed hybrid events, are very precisely measured and
provide an invaluable energy calibration tool.
In order to cover the full sky, the observatory 
will consist of two sites, one in the southern 
hemisphere and one in the north.  The southern site
is located in Mendoza, Argentina, and construction
there is nearing completion, at which time the observatory 
will comprise 24 fluorescence telescopes overlooking 
1600 surface detectors spaced 1.5~km apart on a
hexagonal grid.  Colorado has been selected
as the location for the northern site.

The requirements of this project place rather strong demands
on the software framework underlying data analysis.  Most importantly,
the framework must be flexible and robust 
enough to support the collaborative effort of a large
number of physicists developing a variety of applications over
the projected 20 year lifetime of the experiment. 
Specifically, the software must 
support simulation and reconstruction of events using
surface, fluorescence and hybrid methods, as well as
simulation of calibration techniques
and other ancillary tasks such as data preprocessing.  
Further, as the experimental
run will be long, it is desirable that the software be
extensible in case of future upgrades to the observatory
instrumentation.  The offline framework 
must also handle a number of data formats in order to deal with
event and monitoring information from a variety of instruments, 
as well as the output
of air shower simulation codes.
Additionally, it is desirable that all physics code
be ``exposed'' in the sense that any collaboration member must be able to 
replace existing algorithms with his own in a straightforward
manner.  Finally, while the underlying framework itself may
exploit the full power of C++ and object-oriented design, 
the portions of the code directly used by physicists 
should not assume a particularly detailed knowledge of
these topics.  

The offline framework was designed with these principles 
in mind.  Implementation has taken place over the last
three years, and the first physics results based upon this
code were recently presented at the $29^{\rm th}$ International Cosmic
Ray Conference~\cite{icrc2005}.
  

\section{Overview}

\begin{figure}[htb]
\centering
\includegraphics[width=2.8in]{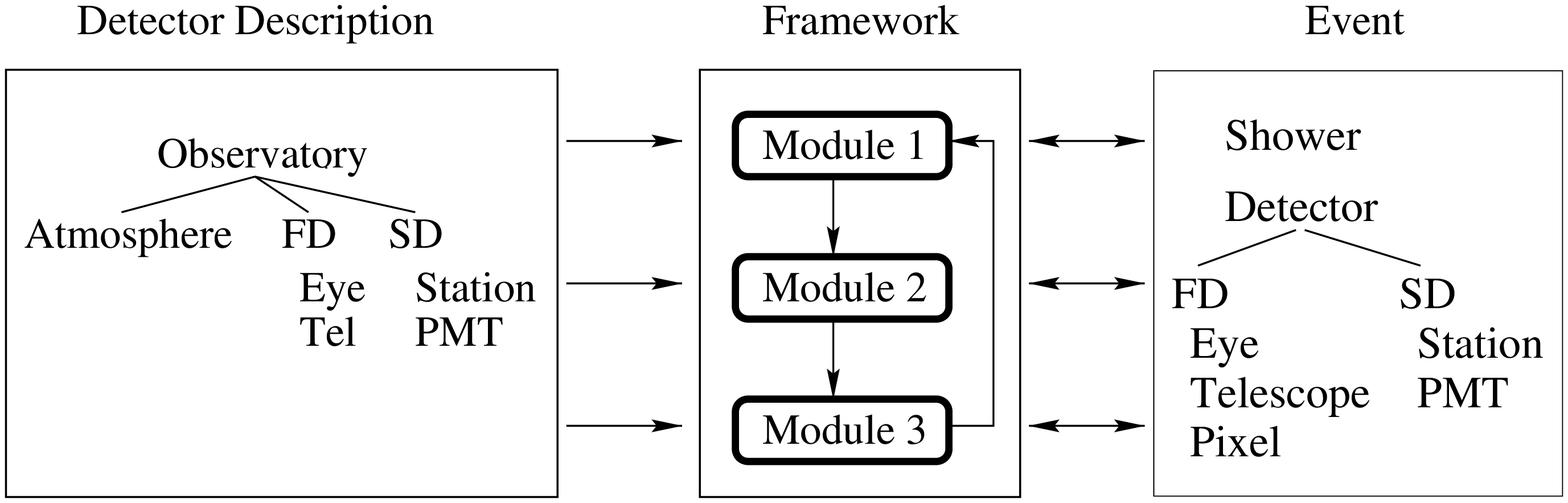}
\caption{General structure of the offline framework.
Simulation and reconstruction tasks are broken down
into modules.  Each module is able to read information
from the detector description and/or the event,
process the information, and write the results back
into the event.}
\label{f:general}
\end{figure}

The offline framework comprises three principal parts:
a collection of processing {\em modules} which can
be assembled and sequenced through instructions 
provided in an XML file, an {\em event} structure
through which modules can relay data to one another
and which accumulates all simulation and reconstruction
information, and a {\em detector description} which provides
a gateway to data describing the configuration and 
performance of the observatory as well as atmospheric 
conditions as a function of time.  These principal 
ingredients are depicted in figure~\ref{f:general}.

These components are complimented by
a set of foundation classes and utilities for
error logging, physics and mathematical manipulation,
as well as a unique package supporting
abstract manipulation of geometrical objects. 

Each of these aspects of the framework is described in more detail
below.

\section{User code, Modules and Configuration} \label{sec:config}

Experience has shown that most tasks of interest of the Pierre Auger
Collaboration can be reasonably factorized into sequences of well-defined
processing steps.  Physicists prepare such processing algorithms
in so-called {\em modules}, which they can insert into the framework
via a registration macro in their code.  This modular design allows
collaborators to easily exchange code, compare algorithms and
build up a wide variety of applications by combining modules in various 
sequences. 

Run-time control over module sequences is afforded through
a {\em run controller} which invokes the various
processing steps within the modules according to a set of externally 
provided instructions.  As most of our applications 
do not require extremely sophisticated sequencing control
at the module level, we have chosen to construct
a very simple XML~\cite{xml}-based language for 
specifying sequencing instructions.  This provides 
users with a tool which can be learned
very quickly, but which is still rich enough to still support most common
applications.  Figure~\ref{f:xml} shows a simple example 
of the structure of a sequencing file.
\begin{figure}[htb]
\centering
\begin{verbatim}
<sequenceFile>
   <loop numTimes="unbounded">
     <module> SimShowerReader </module>
     <loop numTimes="10">
        <module> EventGenerator </module>
        <module> TankSim        </module>
        <module> TriggerSim     </module>
        <module> EventExporter  </module>
     </loop>
   </loop>
</sequenceFile>
\end{verbatim}
\caption{Simplified example in which an XML file
sets a sequence of modules to conduct a simulation
of the surface array.  {\tt <loop>} and {\tt <module>}
tags are interpreted by the Run Controller, which
invokes the modules in the proper sequence.  In this
example, simulated showers are read from a file, and
each shower is thrown onto the array in 10 
random position by an {\tt EventGenerator}.  Subsequent 
modules simulate the response of the surface detectors
and trigger, and export the simulated data to file.}
\label{f:xml}
\end{figure}

Parameters, cuts and configuration instructions used by
modules or by the framework itself are also stored in 
XML files.  A central directory points modules to their configuration
file(s) using a local filename or URI and creates parsers to assist in
reading information from these files.  This directory is
constructed from a {\em bootstrap} file whose name is passed
on the command line at run time.
The configuration
mechanism can also concatenate and write a log file with 
all configuration data accessed during a run.  The log
file format identical to the {\em bootstrap} format, so 
the logs can be subsequently read in
to reproduce a run with an identical configuration.  This 
configuration logging mechanism may also be used to record the versions
of modules and external libraries which are used during a run.

To check configuration files for errors, we employ XML Schema~\cite{schema}
validation throughout.  This has proved successful in saving coding time
for developers and users alike, and facilitates much more detailed
error checking than most users are likely to implement on their own.
All XML handling is based upon the Xerces~\cite{xerces} validating parser,
augmented by a wrapper to simplify use and compliment functionality
with features such as unit handling, expression evaluation, and 
casting of data in XML files to atomic types or STL containers.

\section{Data Access}

The offline framework provides two parallel hierarchies for accessing
data: the {\em event} for reading and writing information that changes per event, 
and the read-only {\em detector description} for retrieving static or slowly
varying information such as detector geometry, calibration constants,
and atmospheric conditions. 

\subsection{Event}

The event data structure contains all raw, calibrated, 
reconstructed and Monte Carlo data and acts as the 
principal backbone for communication between modules.  
As it is a communication backbone, reference semantics are used
throughout to access data structures in the event and constructors are 
kept private to prevent accidental copying of event components.  
The event structure is built up dynamically as needed, and is
instrumented with a simple protocol allowing modules to interrogate the
event at any point to discover its current constituents.  

The event representation in memory is decoupled from the 
representation on disk.  Serialization is currently implemented
using the ROOT~\cite{root} toolkit, though the design is intended
to allow for relatively straightforward changes of 
serialization machinery.  

A set of simple-to-use input/output modules are provided to 
allow users to transfer all or part of the event from memory
to a file at any stage in the processing, and to reload the event and 
continue processing from that point onward.  These modules are 
are built upon a a set of utilities to support the multi-format
reading and writing required to deal with different raw event and
monitoring formats as well as the different formats used by the 
AIRES~\cite{aires}, CORSIKA~\cite{corsika} and CONEX~\cite{conex}
air shower simulation packages.

\subsection{Detector Description} \label{sec:detector}

\begin{figure}[htb]
\centering
\includegraphics[width=2.8in]{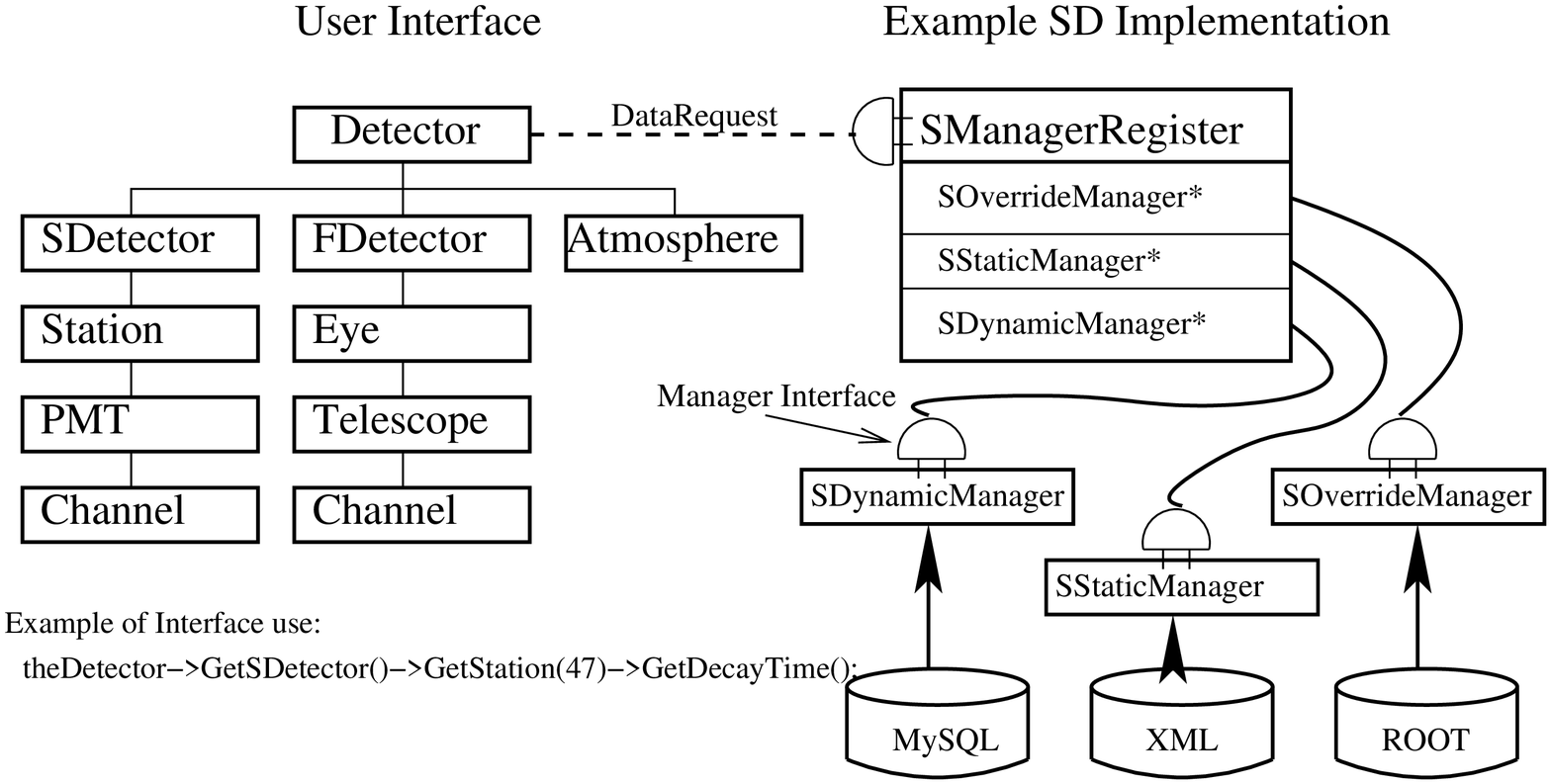}
\caption{
Machinery of the detector description.  The user interface
(left) comprises a hierarchy of objects describing the 
various components of the observatory.  These objects relay
requests for data to registry of managers
(right) which handle multiple data sources and formats.
}
\label{f:detector}
\end{figure}

The {\em detector description} provides a unified interface
from which module authors can retrieve information about
the detector configuration and performance at a particular time.
The interface is organized following the hierarchy normally 
associated with the observatory instruments.  Requests for data
are passed by this interface to a registry
of so-called {\em managers}, each of which is capable of 
extracting a particular sort of information from a 
particular data source.  Data retrieved from the manager
are cached in the interface for future use.  
In this approach, the user deals with a
single interface even though the data sought may reside in 
any number of different sources.  Generally we choose to store
static detector information in XML files, and 
time-varying monitoring and calibration data in MySQL~\cite{mysql}
databases.  The structure of the detector description 
machinery is illustrated in figure~\ref{f:detector}.

Note that it is possible to implement more than one manager
for a particular sort of data.  In this way, a special 
manager can override data from a general manager.  For 
example, a user can decide to use a database for the
majority of the description of the detector, but override some data
by writing them in an XML file which is interpreted by 
the special manager.  The specification of which data 
sources are accessed by the manager registry and in 
what order they are queried is detailed in a configuration
file.  The configuration of the manager registry is transparent
to the user code.

The detector description is also equipped to support a 
set of plug-in functions, called {\em models} which can be used for 
additional processing of data retrieved through the detector.
These are used primarily to interpret atmospheric
monitoring data.  As an example, users can invoke
a model designed to evaluate attenuation due to aerosols 
between two points in the atmosphere.  This model
interrogates the detector interface to find the 
atmospheric conditions at the specified time, and
computes the attenuation.  Models can also be 
placed under command of a {\em super-model} which
can attempt various methods of computing the desired
result, depending on what data are available at the 
specified time.

\section{Utilities}

The offline framework is built on a collection of 
utilities, including a XERCES-based XML parser, an 
error logger, various mathematics and physics services, 
testing utilities and a set of foundation classes to represent
objects such as signal traces, tabulated functions and
particles.  The utilities collection also includes
a novel geometry package, which we describe in more 
detail here.

As discussed previously, the Pierre Auger Observatory
comprises many instruments spread over a large area and, in 
the case of the fluorescence telescopes, oriented in different
directions.  Consequently there is no natural 
preferred coordinate system for the observatory; 
indeed each detector component
has its own preferred system, as do components 
of the event such as the air shower itself.
Furthermore, since the detector spans some 40~km from side to side, 
the curvature of the earth cannot generally be neglected.
In such a circumstance, the necessity of 
keeping track of all the 
the required transformations when performing
geometrical computations is tedious and
error prone.  

This problem is alleviated in
the offline geometry package by 
providing geometrical objects such as points
and vectors which keep track of the coordinate 
system in which they are represented.  
Operations on these objects can then be 
written in an abstract way, 
independent of any particular coordinate system choice.
There is no need for pre-defined coordinate
system conventions, or coordinate system
conversions at module boundaries.

Coordinate systems themselves are defined 
in terms of transformations of other coordinate 
systems, with an ultimate root coordinate system
as the foundation.  In order to avoid
reliance on this root coordinate system by all 
of the client code, a registry of pre-defined 
coordinate systems is provided.
Furthermore, 
various specialized coordinate systems,
such as the coordinate system of the shower
or one of the telescopes, 
can be retrieved from different parts of the 
event and detector description.

Locations of detector components are provided
by survey teams in UTM (Universal Transverse Mercator)
coordinates,  which are convenient for navigation, but less so
for data analysis.
The geometry package therefore includes support for
transformations between geodetic and Cartesian
coordinates.  

\section{Build System and Quality Control}

To help ensure code maintainability and stability in 
the face of a large number of contributors and 
a decades long experimental run, unit and acceptance testing
are integrated into the offline framework build and
distribution system. This sort of quality assurance
mechanism is crucial for any software which must continue
to develop over a timescale of many years.

Our build system is based on the GNU autotools~\cite{autotools},
which provide hooks for integrating tests with the 
build and distribution system.  A substantial collection
of unit tests has been developed, each of which
is designed to comprehensively test a single framework
component.  These unit tests are run at regular
intervals, and in particular prior to releasing a new version
of the software.  We have employed the CppUnit~\cite{cppunit}
testing framework as an aid in implementing these unit tests.
We are currently in the process of developing more involved acceptance
tests which will be used to verify that modules and 
framework components working in concert continue to function
properly during ongoing development.

\section{External packages}

The choice of external packages upon which to build the offline
framework was dictated not only by package features, support
and the requirement of being open-source, but also by our
best assessment of prospects for longevity.  At the same time,
we attempted to avoid locking the design to any single-provider
solution.  To help achieve this, the functionality of 
external libraries is often provided 
to the client code
through wrappers or fa{\c c}ades, as in the case of XML 
parsing described in section~\ref{sec:config}, or through 
a bridge, as in the case of the detector description described
in section~\ref{sec:detector}.  The collection of external
libraries currently employed includes ROOT~\cite{root} for
serialization, Xerces~\cite{xerces} for XML parsing and validation,
CLHEP~\cite{clhep} for expression evaluation and geometry foundations,
Boost~\cite{boost} for its many valuable C++ extensions, and optionally
Geant4~\cite{g4} for detailed detector simulations.

\section{Conclusions}

We have implemented an offline software framework for the Pierre Auger Observatory.
It provides machinery to help collaborators work together
on data analysis problems, compare results, and carry out
production runs of large quantities of simulated or real data.
The framework is configurable enough to adapt to a diverse
set of applications, while the user side remains simple enough
for C++ non-experts to learn in a reasonable time.  The modular
design allows straightforward swapping of algorithms for quick comparisons
of different approaches to a problem.  The interfaces to detector
and event information free the users from having to deal individually
with multiple data formats and data sources.  This software, while still 
undergoing vigorous development and improvement, has been 
used in production of the first physics results from the observatory.

\section*{Acknowledgment}

The authors would like to thank the fearless 
early users of the offline framework.



%




\end{document}